\documentclass[aps,prb,twocolumn,showpacs,superscriptaddress]{revtex4}
\usepackage[latin1]{inputenc}
\usepackage[english]{babel}
\usepackage[T1]{fontenc}
\usepackage{bm,graphicx,graphics,amsmath,amssymb,epsfig,color}

\def \be {\begin{equation}}
\def \ee {\end{equation}}
\def \beA {\begin{eqnarray}}
\def \eeA {\end{eqnarray}}

\def \grad {\vec{\nabla}}
\def \der {\partial}

\def \Re  {\mbox{Re}}

\def \graffb#1{\left\{ #1 \right\}}

\begin{document}

\title{Thermo-magnonic diode: rectification of energy and magnetization currents}

\author{Simone Borlenghi} 
\affiliation{Department of Materials and Nanophysics,  School of Information and Communication Technology, \\Electrum 229, Royal Institute of Technology, SE-16440 Kista, Sweden.}
\author{Stefano Lepri}
\affiliation{Consiglio Nazionale delle Ricerche, Istituto dei Sistemi Complessi,Via Madonna del Piano 10 I-50019 Sesto Fiorentino, Italy.} 
\affiliation{Istituto Nazionale di Fisica Nucleare, Sezione di Firenze, via G. Sansone 1 - I-50019 Sesto Fiorentino, Italy}
\author{Lars Bergqvist} 
\affiliation{Department of Materials and Nanophysics,  School of Information and Communication Technology, \\Electrum 229, Royal Institute of Technology, SE-16440 Kista, Sweden.}
\affiliation{SeRC (Swedish e-Science Research Center), KTH, SE-10044 Stockholm, Sweden.}
\author{Anna Delin}
\affiliation{Department of Materials and Nanophysics,  School of Information and Communication Technology, \\Electrum 229, Royal Institute of Technology, SE-16440 Kista, Sweden.}
\affiliation{SeRC (Swedish e-Science Research Center), KTH, SE-10044 Stockholm, Sweden.}
\affiliation{Department of Physics and Astronomy, Uppsala University, Box 516, SE-75120 Uppsala, Sweden.}

\begin{abstract}
We investigate the dynamics of two coupled macrospins connected to thermal baths at different temperatures. The system behaves like a diode which allows the propagation
of energy and magnetization currents in one direction only. This effect is described by a simple model of two coupled nonlinear oscillators
interacting with two independent reservoirs. It is shown that the rectification phenomenon can be interpreted as a 
a stochastic phase synchronization of the two spin-oscillators. A brief comparison with realistic micromagnetic simulations 
is presented. This new effect yields promising opportunities in spin caloritronics and nanophononics devices. 
\end{abstract}

\maketitle

\section{Introduction}

Since the discovery of the spin-Seebeck effect \cite{uchida08,sinova10}, according to which a thermal gradient in a ferromagnet generates a spin current, 
the emerging field of spin-caloritronics \cite{bauer11} has been the object of intense investigations. 

A related line of research, that was developed independently in the recent years, focuses on heat transport in lattices of 
nonlinear oscillators \cite{lepri03,DHAR09}. The relevance of such studies to condensed-matter problems is testifyied 
by the growing interest for heat transport properties of low-dimensional materials like nanotubes \cite{Chang08} 
or graphene \cite{balandin12}. Further motivations come from the perspective of controlling nanoscale 
energy flows \cite{terraneo02,casati04} as well as from the hope of finding novel 
dynamical mechanisms that could enhance the efficiency of thermoelectric energy conversion~\cite{Saito2010}.

Within this general background, in the present work, we investigate theoretically a system that could be the building block of novel magnonic devices, 
allowing the propagation of energy and magnetization currents in one direction only. The system consists of two coupled macrospins connected to thermal reservoirs at different temperatures. 
It has been recently demonstrated by means of micromagnetic simulations that such a system can indeed act a spin-Seebeck diode \cite{borlenghi13}.  
The basic functioning principle is similar to the one of the thermal diode considered in the recently born field of 
phononics \cite{terraneo02,casati04,chang07,balandin12,li12}. It can be qualitatively explained in terms of a temperature-dependent
renormalization of the macrospins frequency spectra whose overlap may lead to a conducting or nonconducting state
depending on the sign of the applied thermal gradient. However, such thermo-magnonic system offers several new possibilities
for control of nanoscale energy flows \cite{ren13}. The most evident one is related to the fact that we are here dealing with \emph{two coupled currents}
of the basic conserved quantities, energy and magnetization.

To get a theoretical insight we study here an effective simplified model consisting of two coupled oscillators interacting with 
external reservoirs. This will allow to put on a more clear basis the basic operating principles of the system. 
In particular, we will argue that the rectification effect can be described as  
\emph{stochastic phase synchronization} (SPS) \cite{stra1967,neim1998,neiman99,teramae04} of the two precessing spins. 
SPS occurs in a large class of nonlinear oscillator driven by noise. It basically amounts to the 
fact that noise can lead to an enhanced phase entrainment and thus to an increase of the energy transfer
among the oscilllators. This phenomenon, which has attracted large interest in the past decade in connection with 
biomedical systems and neural circuits \cite{singer99,bahar02},
has never been investigated in the broad context of nanoscale energy transfer and in particular in magnonics and spin-caloritronics. 
Such novel interpretation, that will be pursued in the present work, is useful to ease the physical intuition and to 
suggest a new mechanism for the transfer of energy and spin currents in those systems.

The present study answer a very general question, that is, under which condition the transfer of energy an magnetization between coupled spins at different temperatures occurs.
The rectification effect considered here has several applications. In particular, it opens the way to the experimental realization of thermal logic gates, which have been recently described theoretically within the field of 
nanophononics \cite{kim10,kruglyak10,khitun10}. This could be the starting point of a new generation of energy efficient electronic devices.

Moreover, we suggests to implement this new diode using a very common and well known spintronics device, the spin valve.
We wish to point that, the studies on the spin-Seebeck effect on spin-valve systems performed so far, concern only the spin current carried by electrons.
The notion of magnetization current developed here, which is a special case of the usual spin-wave current, has not been investigated so far in this kind of systems.
Then, spintronics experiments are challenging for the intrinsic difficulty of measuring the spin current, which is based on the inverse spin-Hall effect \cite{uchida08}.
Here, we suggest a \emph{direct} way to detect spin and energy transport, which is based on the overlap of the SW modes of the system. This can be done using various well known techniques, such as
the ferromagnetic resonance force spectroscopy \cite{thesis,naletov11}.

The paper is organized as follows. In Sec.~\ref{sec:model} we describe the physical system and introduce the 
effective coupled-oscillator model and its interaction with the thermal reservoirs. In Sec.~\ref{sec:phase} we 
discuss the rectification of energy and magnetization currents as a phase-synchronization phenomenon induced
by the thermal fluctuations. Some simulations of the oscillators model are thereby described. In Sec.~\ref{sec:coupled}
we consider the case in which the system is driven by both a thermal and magnetization gradient.
The predictions of the models are then tested qualitatively with the micromagnetic simulations of the 
full magnetization dynamics of the device (Sec.~\ref{sec:micro}). Finally, we close the paper with 
some overview in the concluding section.

\section{Physical system and model}
\label{sec:model}

\begin{figure}
\begin{center}
\includegraphics[width=8cm]{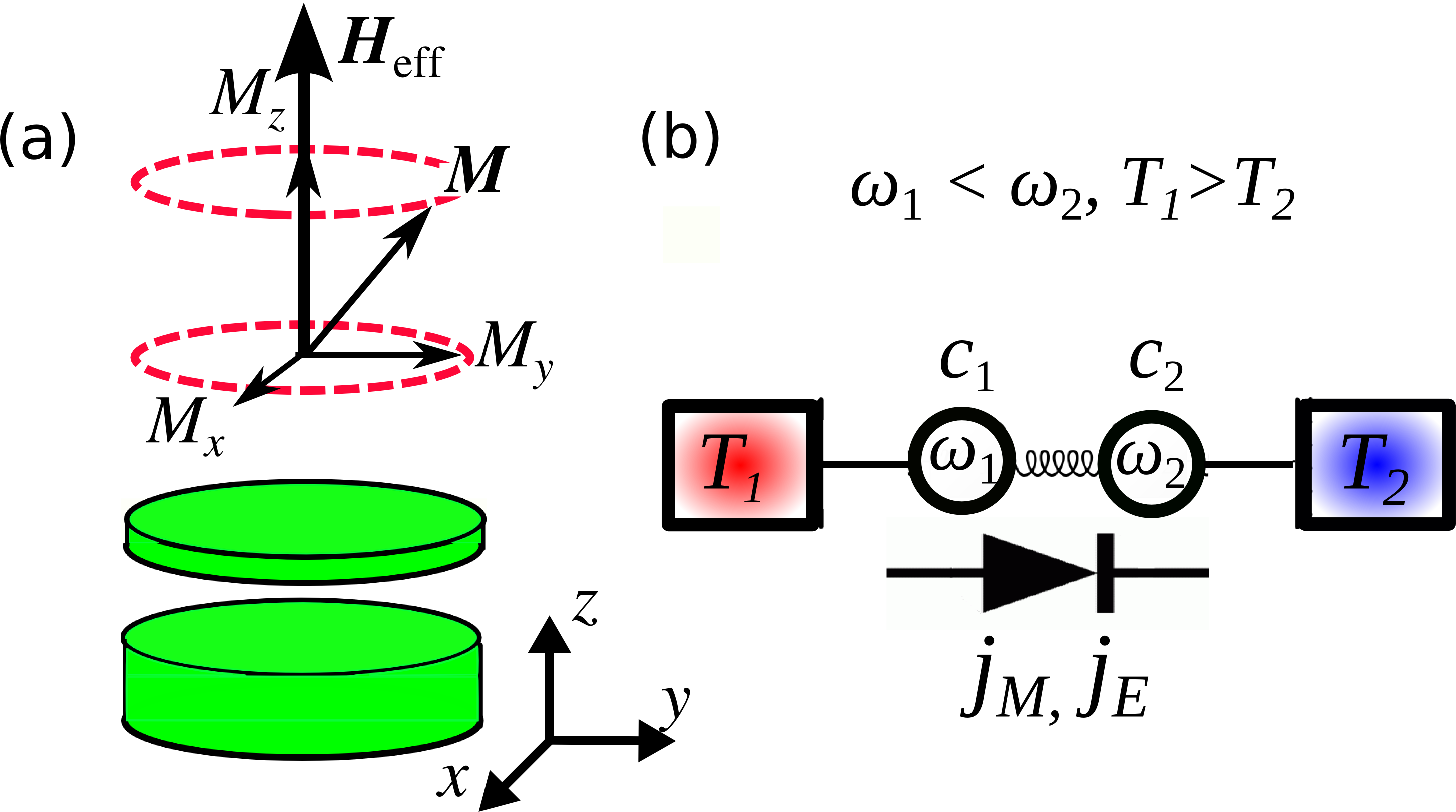}
\end{center}
\caption{(Color online) (a) circular precession of the magnetization in a system of two disks coupled via dipolar interaction. 
(b) The system behaves as chain of two oscillators with frequencies $\omega_1<\omega_2$ connected to two thermal baths.
\emph{Rectification effect}: When $\Delta T=T_1-T_2>0$ the two frequencies may overlap, 
giving the two non-zero currents $j_M$ and $j_E$. When $\Delta T<0$ the frequencies split, 
and no net current flows in the system.}
\label{fig:system}
\end{figure}

The local dynamics of the magnetization $\bm{M}$ in a ferromagnet is described by the Landau-Lifshiz-Gilbert (LLG) equation
\cite{landau65,gilbert55,gurevich96}
\be\label{eq:llg}
\frac{\der{\bm{M}}}{\der t} = -\gamma_0\bm{M}\times\bm{H}_{\rm{eff}}+\alpha{\bm{M}}\times\frac{\der{\bm{M}}}{\der t}/M_s
\ee
where $\gamma_0$ is the gyromagnetic ratio, $\alpha$ is the adimensional Gilbert damping parameters and $\bm{H}_{\rm{eff}}$ is the effective field, given by the functional
derivative of the Gibbs free energy of the system with respect to the magnetization. In our case, the effective field contains the applied, exchange and dipolar fields. 
The first term at the rhs of Eq.(\ref{eq:llg}) describes the precession of the magnetization around the effective field with frequency $\omega=\gamma_0|\bm{H}_{\rm{eff}}|$, 
while the second term accounts for energy losses at a rate proportional to the Gilbert Damping parameter $\alpha$.

It is known that an electrical current with spin polarization $\bm{p}$ exerts a spin transfer torque (STT) on the localised magnetic moments of the ferromagnet \cite{slonczewski96,berger96}.
The effect of STT is described by rescaling the effective field as $\bm{H}_{\rm{eff}}\rightarrow\bm{H}_{\rm{eff}}-a\bm{p}$ and by adding to the right hand side of Eq.(\ref{eq:llg}) the term
\be\label{eq:st}
\bm{\tau}=\gamma_0\frac{b}{M_s}{\bm{M}}\times({\bm{M}}\times{\bm{p}}).
\ee
The terms proportional to $a$ is usually called field like component of STT. The term proportional to $b$, which controls the damping, is the usual Slonczewski STT \cite{hitoshi07,sankey08,petit08,jung10,jia11}. 
The latter can lead to a steady state precession and to a reversal of the magnetization \cite{katine00,katine00b,katine08}. 

The parameters $(a,b)$ are proportional to the intensity of the current and to the degree of spin-polarization. In general, they both depend on the geometry of the system and on the microscopic transport properties
of the material \cite{brataas06}. For their computation in realistic devices, several methods have been developed \cite{brataas00,waintal00,rychkov08,borlenghi11}. Here, they are considered as free parameters of the model, that can be used to control the rectification effect.

The device considered here, shown in Fig.~\ref{fig:system}a, consists of a spin-valve nanopillar made of two ferromagnetic layers separated by a non-magnetic spacer and coupled by dipolar 
interaction. This system, which is the prototype for spintronic devices, has several applications \cite{naletov11,slavin09} and constitutes the usual geometry for spin transfer nano oscillators (STNOs).
 
We consider the simple case where both the effective field and the polarization vectors are aligned with the $z$ axis, which defines the precession axis of the magnetization. 
In this case, the non adiabatic STT leads simply to a rescaling of the oscillation frequency. Thus, in our model the relevant physics is described only by the Slonczewski STT. 
This simplification is quite realistic if we consider the nano-pillar geometry, where it has provided to theoretical descriptions in good agreement with experiments \cite{petit08,naletov11,thesis}.

Whithin the macrospin approximation, the circular precession in the $x$-$y$ plane of the magnetization vectors of the two disks can be described by two coupled LLG equations. 
In the weakly nonlinear regime, those can be effectively approximated in terms of the 
complex spin wave (SW) amplitudes 
\be
c_n=\frac{M_{xn}+iM_{yn}}{\sqrt{M_{sn}(M_{sn}+M_{zn})}} 
\ee
of disk $n=1,2$  \cite{slavin09,naletov11} . At this level of description, the system dynamics can be thus modeled
by the stochastic equations
\beA
\dot{c}_1 & = & (i-\beta_1)(\omega_1c_1+2p_1-h_{12}c_{2})+\sqrt{D_1}\xi_1\label{eq:osc1},\\
\dot{c}_2 & = & (i-\beta_2)(\omega_2c_2+2p_2-h_{21}c_{1})+\sqrt{D_2}\xi_2\label{eq:osc2}.
\eeA
Those are the equation of motion of two coupled nonlinear oscillators, whose resonance frequencies
$\omega_{n}(p_n)\propto |\bm{H}_{\rm{eff}}|$ depend on the SW power $p_n=|c_n|^2$ ($n=1,2$). The analytical expressions for $\omega_n$ at zero temperature,
obtained diagonalizing Eqs.(\ref{eq:osc1}) and (\ref{eq:osc2}), are given in Refs.[\onlinecite{naletov11,thesis}]. 
The damping rates, which describe energy dissipation towards the environment, are chosen to be $\Gamma_n(p_n)=\beta_n\omega_n(p_n)$\cite{slavin09}.
The parameters $\beta_n$ here model the effect of STT.
 
Thermal fluctuations are accounted by the stochastic terms $\sqrt{D_n}\xi_n$, $\xi_n$ being complex Gaussian random variables 
with unit variance and zero average, and $D_n=2\alpha k_BT_{n}$, as prescribed by the fluctuation-dissipation theorem \cite{slavin09}. 
This is equivalent to adding a fluctuating therm to the effective
field in Eq.(\ref{eq:llg}). 

The coupling term $h_{12}c_{2}$ \cite{naletov11} is the functional derivative $i\delta \mathcal{H}_{\rm{int}}/\delta c_{n}^*$ of the interaction Hamiltonian 
$\mathcal{H}_{\rm{int}}=h_{12} c_{1} c_{2}^*+\mbox{c.c}$. Notice that this Hamiltonian has a general form that describes also exchange interaction, magnon tunneling 
between different material and phase locking in STNOs arrays \cite{katine08,slavin09}. 

The chosen form of the stochastic and dissipative terms insures that, for $\beta_n=\alpha$ and $T_1=T_2=T$, the expected canonical distribution $\exp\graffb{-\mathcal{H}/(k_{B}T)}$
is the stationary solution of the the Fokker-Planck equation associated to Eqs.(\ref{eq:osc1}) and (\ref{eq:osc2})  
(here $\mathcal{H}$ is the Hamiltonian for the isolated system \cite{iubini13}). In this first part of the paper we will focus on the case in which $\beta_n=\alpha$. 
The case when $\beta_n\neq\alpha$, where the system is kept out of equilibrium by STT \cite{slavin07}, is also of interest and will be discussed in Sec \ref{sec:coupled}. 

It is important to remark that Eqs.(\ref{eq:osc1}) and (\ref{eq:osc2}) hold if the system is dominated by two SW modes, one for each oscillator. In the presence of a texture of the magnetization,
the system contains several SW modes, which depend essentially on the geometry of the system \cite{naletov11,thesis}. This feature can be described developing Eqs. (\ref{eq:osc1}) and (\ref{eq:osc2}) 
in the proper SW mode basis, as in Refs.[\onlinecite{naletov11,thesis}]. 
In these systems, thermal fluctuation excite all the SW modes of the system and the intrinsic nonlinearity of the LLG equation generates an additional coupling between the SW modes. 
The latter can originate complex phenomena, such as mode hopping or mode coexistence \cite{muduli12,heinonen13}. In many situations, when the dwelling time between different modes is 
large enough, one can still use Eqs.(\ref{eq:osc1}) and (\ref{eq:osc2}) \cite{muduli12}. 
In our case, it is possible to identify clearly the modes that belong to each oscillator, and the micromagnetic simulations reported in Sec.V corroborate the single mode picture. In perpendicularly magnetized spin valve 
nano-pillars with several SW modes, it has been shown that \cite{thesis,naletov11,borlenghi13} a SW mode expansion of Eqs.(\ref{eq:osc1}) and (\ref{eq:osc2}), 
is sufficient to describe the system properly even in the presence of thermal fluctuations. In particular, no mode hopping as been observed, even in the presence of thermal gradient \cite{borlenghi13}.

Let us now introduce the conserved currents of the system.
Combining Eqs.(\ref{eq:osc1}) and (\ref{eq:osc2}) with their complex conjugates gives the two conservation equations for the SW power \cite{slavin09}
\beA
\dot{p}_1 & = & -2\Gamma_1(p_1) p_1 +j^{12}_M,\label{eq:consmag1}\\
\dot{p}_2 & = & -2\Gamma_2(p_2) p_2 +j^{21}_M,\label{eq:consmag2}
\eeA
which leads to the definition of the magnetization current in between the two oscillators \cite{iubini12,iubini13}: 

\be\label{eq:magcur}
j^{12}_M=2h_{12}\mbox{Im}(c_1c_{2}^*). 
\ee
Notice that Eqs.(\ref{eq:consmag1}) and (\ref{eq:consmag2}) are the conservation equations for the $z$ component of the magnetization. For a continuum ferromagnet with exchange stiffness $A$, they
leads to the usual definition of SW spin current $\bm{j}_{M}=A\bm{M}\times\grad\bm{M}$ carried by the exchange interaction \cite{kim10}.
The conservation equation for the local energy gives the energy current

\be\label{eq:ecur}
j^{12}_{E}=2h_{12}\Re({\dot{c}_1c_{2}^*}),
\ee
which describes the transfer of energy between the oscillators. The computation of those currents is similar to the case of the discrete nonlinear Schroedinger equation, see Refs.[\onlinecite{iubini12,iubini13}]
for a thorough discussion.

\section{Phase dynamics and rectification}
\label{sec:phase}

Let us now discuss the rectification effect. A full analytical solution is obtained in principle solving the Fokker-Planck equation associated to Eqs.(\ref{eq:osc1}) and (\ref{eq:osc2}), 
in a way similar to Ref.[\onlinecite{liu13}]. For our purposes, it suffices to restrict first to a discussion of the deterministic equations, 
obtained by sample-averaging Eqs.(\ref{eq:osc1}) and (\ref{eq:osc2}).
Setting $c_n=\sqrt{p_n}\mbox{e}^{i\theta_n}$ and $\phi=\theta_1-\theta_2$, those equations are written in the phase-amplitude representation as
\beA
\dot{p}_1 & = & -2\Gamma_1(p_1)p_1-j^{12}_M+2D_1,\label{eq:osc_det1}\\
\dot{p}_2 & = & -2\Gamma_2(p_2)p_2+j^{21}_M+2D_2,\label{eq:osc_det2}\\
\dot{\phi} & = & \omega_1(p_1)-\omega_2(p_2)+\label{eq:osc_phi}\\
& + & (h_{21}\sqrt{p_1/p_2}-h_{12}\sqrt{p_2/p_1})\cos{\phi}\nonumber,
%\dot{\phi} & = & \omega_1(p_1)-\omega_2(p_2)+\roundb{h_{21}\sqrt{\frac{p_1}{p_2}}-h_{12}\sqrt{\frac{p_2}{p_1}}}\cos{\phi}\label{eq:osc_phi},
\eeA
where the currents read $j^{12}_{M}=2h_{12}\sqrt{p_1p_2}\sin\phi$ and $j^{12}_{E}=2h_{12}\omega_1(p_1)\sqrt{p_1p_2}\sin{\phi}$.
The constant terms $2D_n$ account for the fact that the powers are always bounded away from zero due to fluctuations. 
In the context of phase-synchronization phenomena, Eq.(\ref{eq:osc_phi}) is often referred to 
as the Adler equation \cite{stra1967,neiman99}.  

The solutions of Eqs.(\ref{eq:osc_det1}-\ref{eq:osc_phi}) are of two types: (i) phase running (desynchronized) solutions, where the two oscillators have different frequencies
and (ii) phase-locked (synchronized) ones. 
In case (i) the time-averaged currents are zero, and Eqs.(\ref{eq:osc_det1}-\ref{eq:osc_phi}) reduce to $\Gamma_n(p_n) = D_n$ and $\dot{\phi} = \omega_1(p_1)-\omega_2(p_2)$,
which implies the equipartition relation $p_n=k_BT_n/\omega_n(p_n)$, for $n=1,2$. This means that the thermostats thermalize each oscillator independently and
there is \emph{no net transfer} of energy and magnetization between the oscillators. Notice that the mere fact that $p_1\neq p_2$ \emph{does not} imply that there is a net current: 
the average energy provided by the baths is returned to them. For case (ii) there is instead 
a common frequency of oscillation $\dot{\theta}_n=\omega$, so that $\dot{\phi}=0$ and one has
\beA
\Gamma_1(p_1)&  = &   D_1-j^{12}_{M}/2\label{eq:gamma1},\\
\Gamma_2(p_2)&  = &   D_2+j^{21}_{M}/2\label{eq:gamma2},\\
\omega_1(p_1)& - & \omega_2(p_2) + \label{eq:adler}\\
&+&(h_{21}\sqrt{p_1/p_2}-h_{12}\sqrt{p_2/p_1})\cos\phi=0.\nonumber
\eeA
Such phase-locked regime can only occur if Eqs.(\ref{eq:gamma1}-\ref{eq:adler}) admit a solution, namely for
\be\label{eq:synchro}
|\omega_1(p_1)-\omega_2(p_2)|\leq|h_{21}\sqrt{p_1/p_2}-h_{12}\sqrt{p_2/p_1}|.
\ee
It has to be remarked that, already at this 
level of approximation, all the parameters 
are temperature dependent since the spin-powers $p_1$ and $p_2$, solutions of Eqs.(\ref{eq:gamma1}-\ref{eq:adler}), depend on both $T_1$ and $T_2$.

The crucial observation is that
Eqs.(\ref{eq:gamma1}) and (\ref{eq:gamma2}) are not invariant with respect to the 
exchange of the two noise sources $D_1$ and $D_2$, so there may be regions of the parameters $(D_n,\Gamma_n(p_n))$ where the currents are different
upon exchanging the sign of the applied temperature gradient. In particular, there may be cases in which Eq.~(\ref{eq:synchro}) is satisfied
for, say, $T_1>T_2$ but \textit{not for} $T_1<T_2$ thus yielding the desired effect.
Notice also that Eq.~(\ref{eq:synchro}) defines the condition for the approximate resonance of the 
effective (temperature-dependent) frequencies and is thus conceptually similar to the criteria 
of spectral overlap usually invoked to explain the working principle of phononic thermal diodes 
\cite{terraneo02,casati04}. 

In the previous analysis, thermal noises are accounted for only through their mean values.
In presence of noise it is known that the phase-locking is only effective as fluctuations will eventually
desynchronize the oscillators \cite{stra1967,neim1998}.
In other words even when condition (\ref{eq:synchro}) holds the phases will not remain exactly locked but will undergo
random phase-slips leading to phase diffusion.

\begin{figure}
\begin{center}
\includegraphics[width=5.6cm]{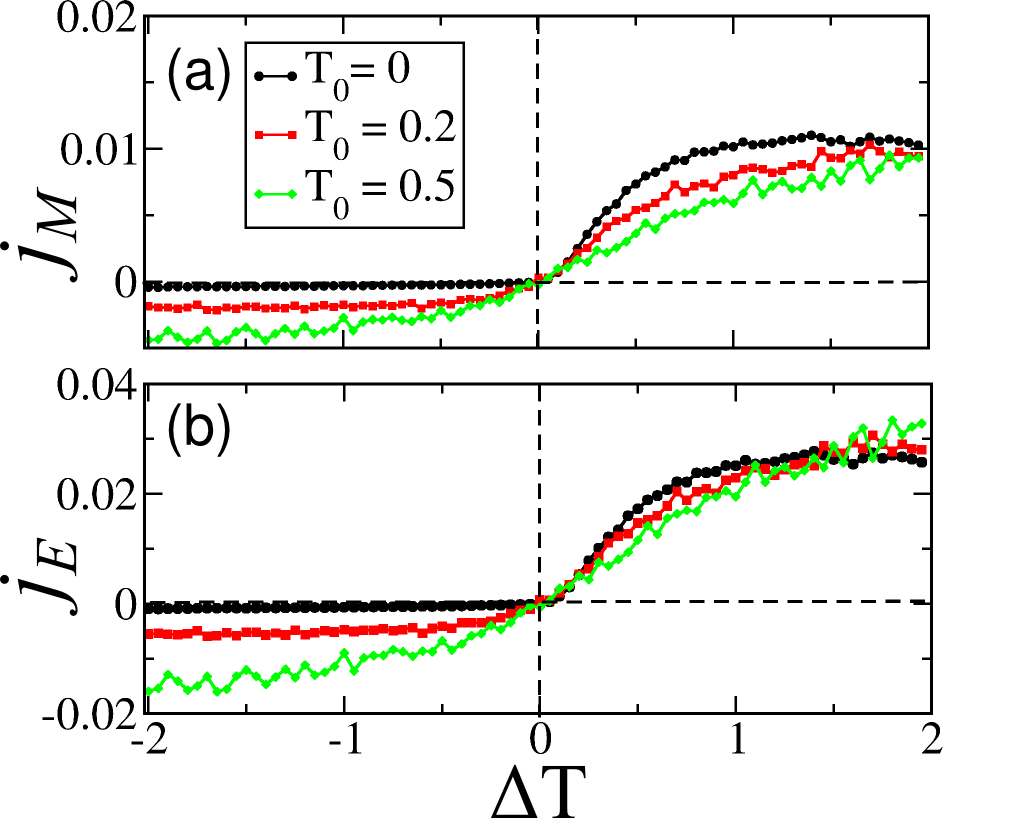}
\end{center}
\caption{(Color online) Rectification effect for the magnetization (a) and energy (b) currents, computed for different values of $T_0$.
Eqs. (\ref{eq:osc1}) and (\ref{eq:osc2}) were integrated numerically using a fourth order Runge-Kutta method with time step $10^{-3}$ model units, 
frequencies $\omega_1=1$ and $\omega_2=2$ and $k_B=1$. }
\label{fig:rectification}
\end{figure}

To substantiate the above arguments, we turn now to numerical simulations of Eqs. (\ref{eq:osc1}) and (\ref{eq:osc2}).
For simplicity, we have taken a symmetric coupling $h_{12}=h_{21}=h$. At equilibrium, we have set $T_1=T_2=T_0$ and then we have increased one of the two temperatures at a time, keeping 
the other fixed at $T_0$ and defining the temperature difference as $\Delta T=T_1-T_2$.
After the system has reached a stationary state, the currents were time-averaged over an interval of $10^6$ time steps.

Figure (\ref{fig:rectification}) shows the two currents vs $\Delta T$, for $\beta_n=\alpha=0.02$ and $h=0.1$, averaged over 50 samples.  The system clearly displays a 
rectification effect when $\Delta T>0$. The two currents have a similar profile, growing 
monotonically until they reach a plateu at $\Delta T\approx 1.2$. Notice that the strength of the rectification effect decreases as $T_0$ increases. 
reduced increasing $T_0$.

\begin{figure}
\begin{center}
\includegraphics[width=6cm]{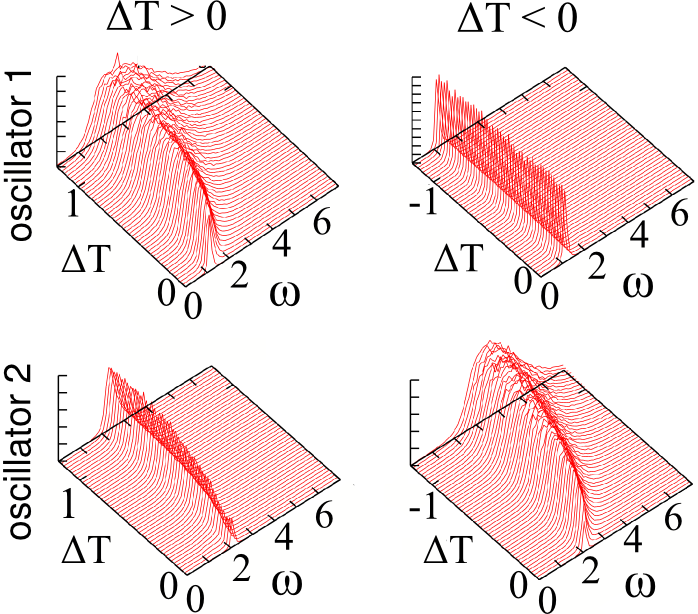}
\end{center}
\caption{(Color online) Power spectra of the two oscillators illustrating the mechanism of resonance underlying the rectification effect.}
\label{fig:spectra}
\end{figure}

The origin of the rectification is illustrated in Figs.(\ref{fig:spectra}), which show the power spectra averaged over 500 trajectories. 
All the following simulations were performed with $T_0=0.2$. 
For positive gradients, the peak at $\omega_1$ broadens and shifts towards higher frequency, until it overlaps with the peak at $\omega_2$, while for
negative gradient the peaks do not overlap. 

\begin{figure}
\begin{center}
\includegraphics[width=6.3cm]{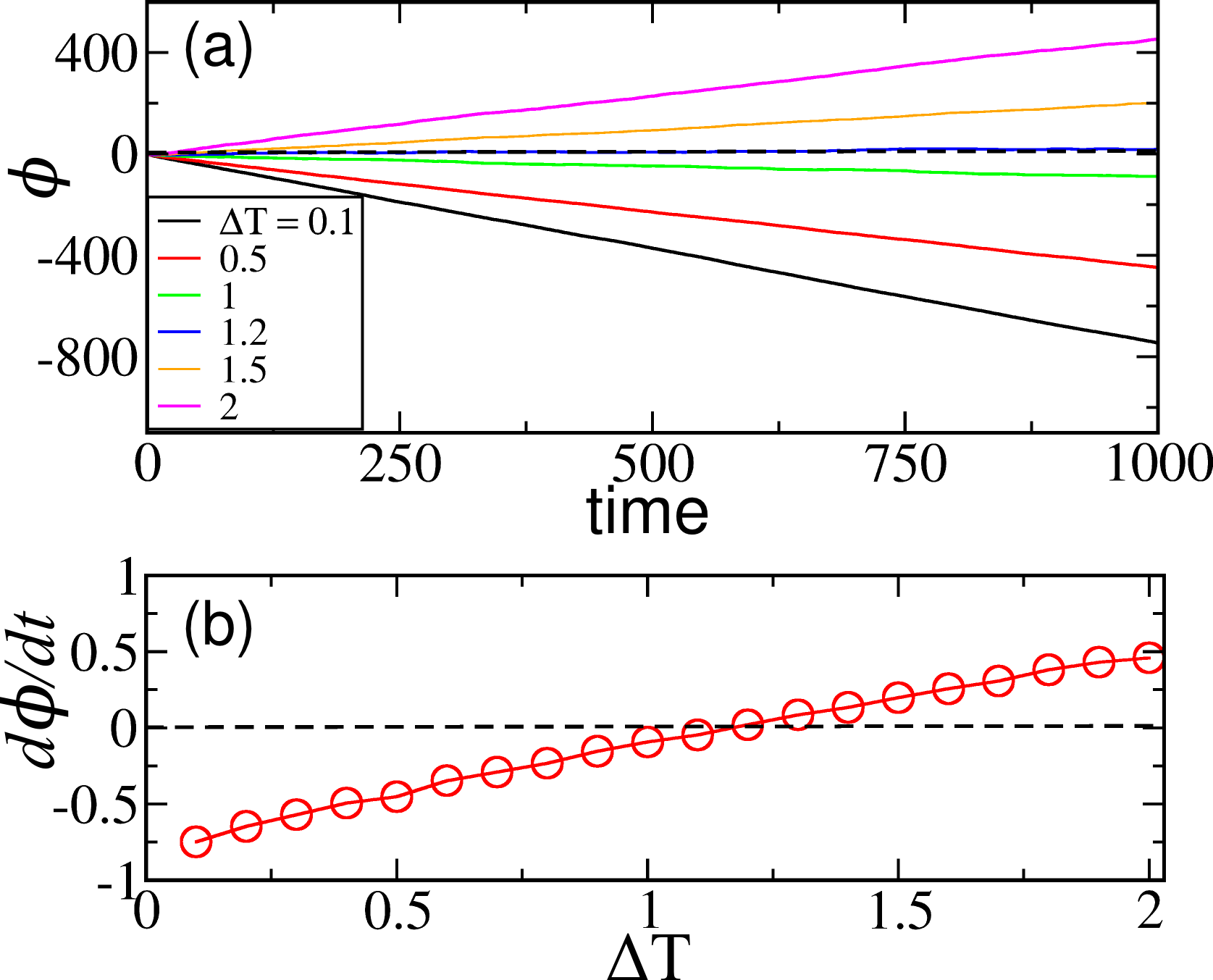}
\end{center}
\caption{(Color online) (a) phase difference $\phi$ vs time, computed for different values of $\Delta T$. $\phi$ increases linearly in time in the desynchronized regime, while it is constant in the synchronized one.
(b) Slope of $\phi$ vs $\Delta T$, which vanishes when the oscillators are synchronized. The line is a guide to the eye.}
\label{fig:phases}
\end{figure}

When the oscillators are phase locked, the time-averaged currents are not zero and there is a \emph{net transport} of energy and magnetization through the system. 
The phase locking can be seen in Fig.(\ref{fig:phases}a), which shows the phase difference $\phi=\theta_1-\theta_2$ vs time, computed for different values
of $\Delta T$ and averaged over 150 samples. One can see that $\phi$ is constant in the synchronized regime. The slope $d\phi/dt$, displayed in Fig.(\ref{fig:phases}b), increases linearly with $\Delta T$, 
and intercepts zero at $\Delta T=1.2$, where the oscillators are synchronized and the currents reach the plateau shown in Fig.(\ref{fig:rectification}).

\begin{figure}
\begin{center}
\includegraphics[width=6cm]{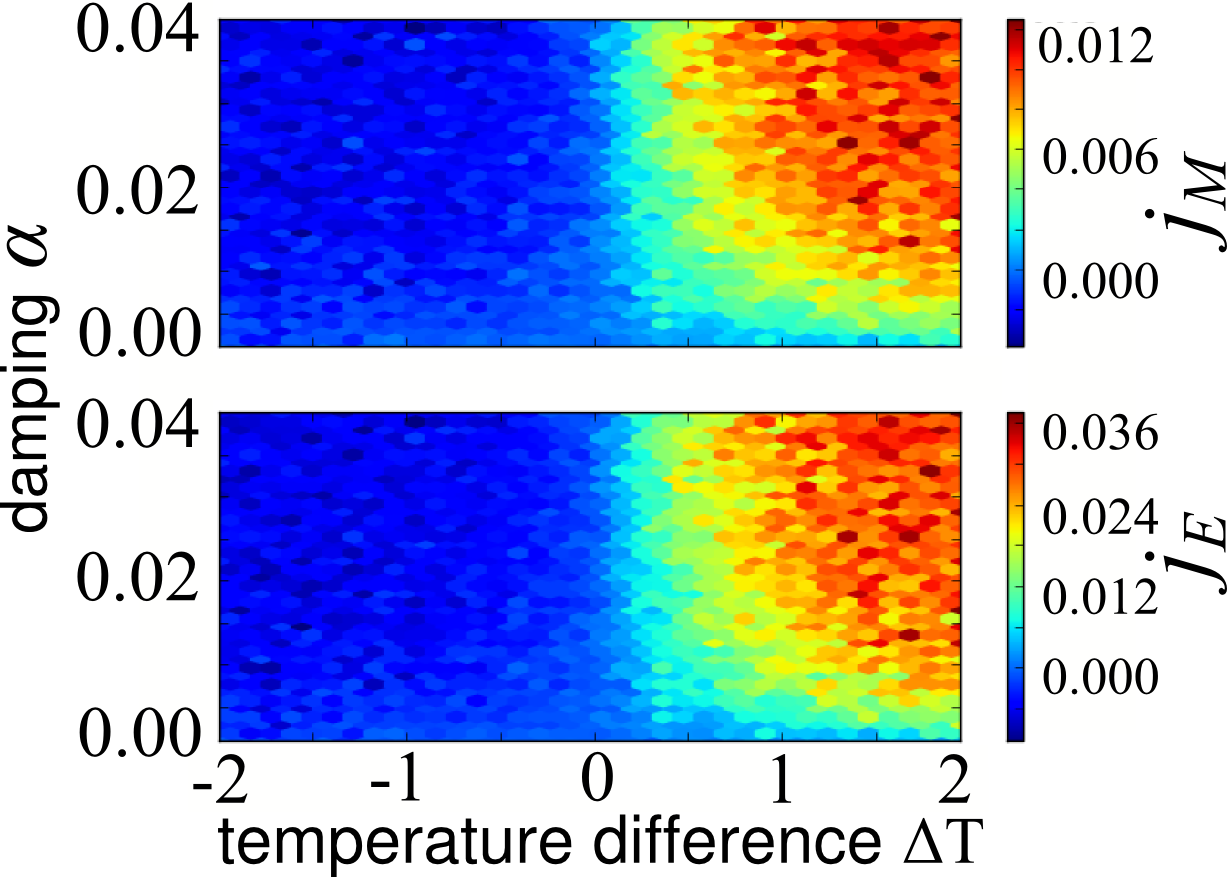}
\end{center}
\caption{(Color online) Magnetization and energy currents as a function of temperature difference $\Delta T$ and Gilbert damping parameter $\alpha$. The currents vanish when the coupling
with the bath $\alpha\rightarrow 0$.
The data were obtained averaging the currents over $2\times10^6$ time steps, with only one trajectory.}
\label{fig:current_alpha}
\end{figure}

\begin{figure}
\begin{center}
\includegraphics[width=6cm]{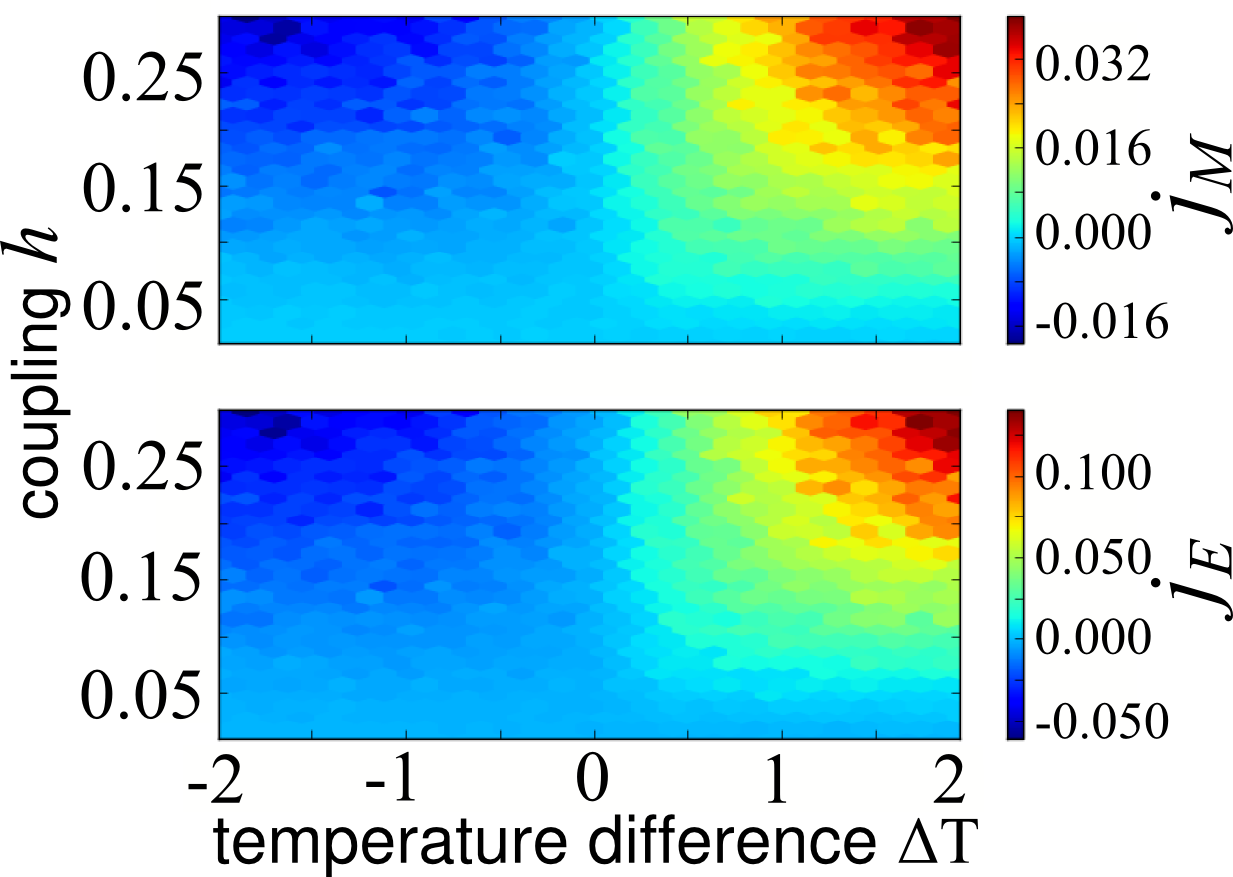}
\end{center}
\caption{(Color online) Magnetization and energy currents as a function of temperature difference  $\Delta T$ 
and coupling strength $h$.  The currents vanish when $h\rightarrow 0$, and the oscillators become uncoupled.
The data were obtained averaging the currents over $2\times10^6$ time steps, with only one trajectory.}
\label{fig:current_coupling}
\end{figure}

We have also investigated the dependence of the current on the damping $\alpha$ and coupling $h$. 
Figures (\ref{fig:current_alpha}) and (\ref{fig:current_coupling}) show the phase diagrams of the currents 
in the planes $(\alpha, \Delta T)$ and $(h,\Delta T)$ respectively. 
Interestingly, the rectification effect is present in a wide range of system parameters.
In both cases, the currents increase with the parameters $\alpha$ and $h$, 
and vanish around $\alpha\approx 10 ^{-3}$ and $h\approx 5\times10^{-2}$. 
This feature depends on the fact that $\alpha$ and $h$ control respectively the coupling with the 
thermal baths and between the oscillators.

\begin{figure}
\begin{center}
\includegraphics[width=8.2cm]{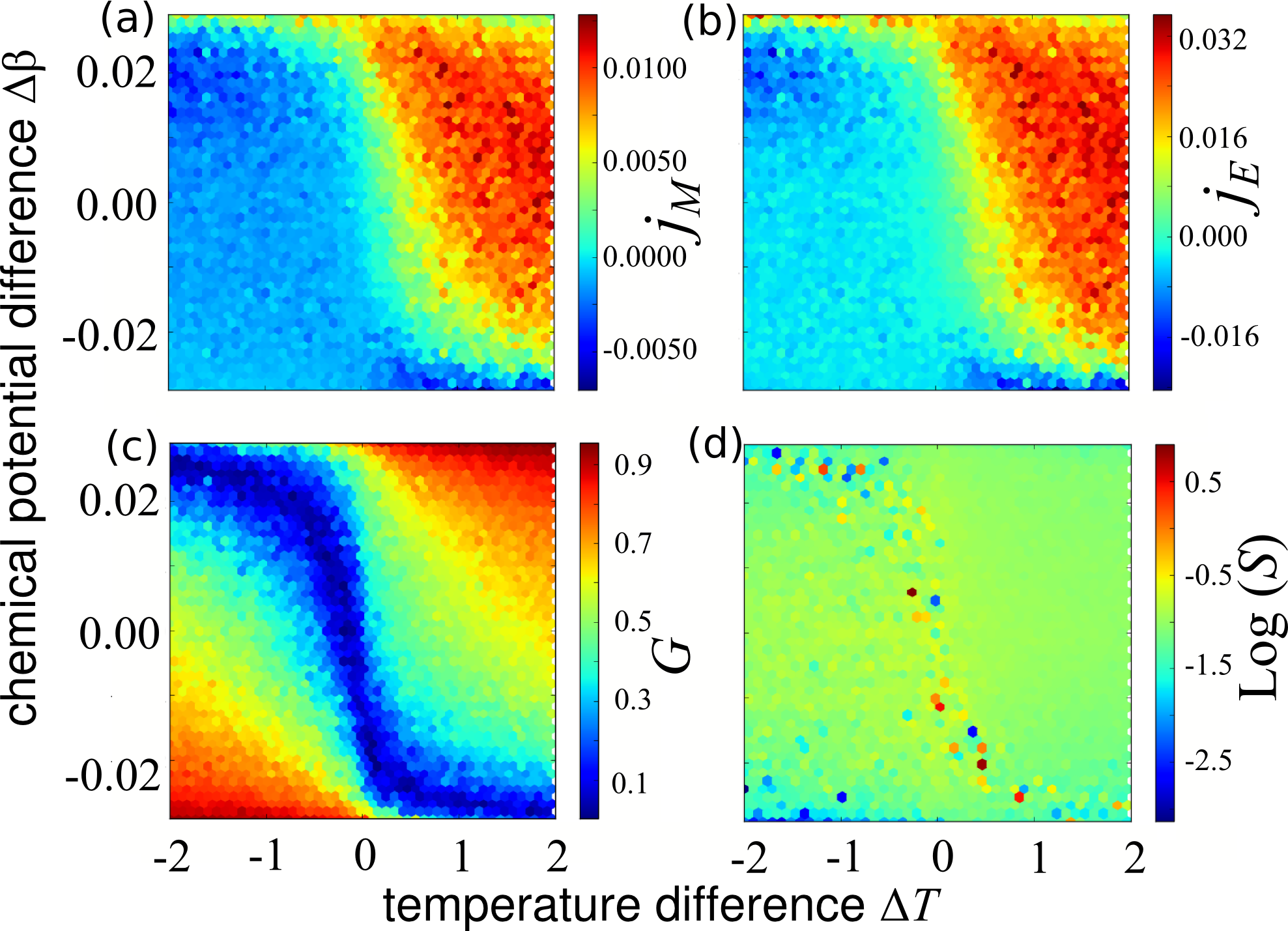}
\end{center}
\caption{(Color online). Phase diagrams in the $(\Delta T,\Delta\beta)$ plane. Panels a) and b) show respectively the magnetization and energy currents, while 
c) and d) display respectively the ratio between the SW powers and the ratio between magnetization and energy current.}
\label{fig:main_result}
\end{figure}

\section{Coupled transport}
\label{sec:coupled}

Up to now we have considered the case in which the damping coefficients $\beta_n$ are set to be equal to the Gilbert damping parameter 
$\alpha$. Actually, in STNOs, the damping can be modified by a spin-polarized current\cite{katine08,slavin09,naletov11}, 
an effect that can be modeled changing the parameter $\beta_n$. This simple fact immediately suggests another route
to drive the system off equilibrium. 
It should in fact be realized that setting $\beta_1 \neq \beta_2$ is somehow equivalent
to apply an external force capable to drive energy and magnetization flows. The situation is analogous 
to the standard non-equilibrium thermodynamics, where 
the two coupled currents currents are associated to two "forces": the differences of temperature and of chemical potential. 
In our system, the parameters $\beta_n$ control the escape rate of the magnons towards the reservoirs \cite{slavin09,naletov11} and 
$\Delta \beta= \beta_2-\beta_1 $ acts as an additional force
that controls the currents, in a way similar to a chemical potential \cite{iubini12,iubini13}. 

Taking $h=0.1$ and $\alpha=0.02$, the dynamics was computed for different values of $\Delta T$ and of 
"chemical potential" difference $\Delta \beta$.
The computations were performed starting at equilibrium with $\beta_n=\beta_0=0.03$, and decreasing one damping while keeping the other fixed at $\beta_0$. 

The phase diagrams of the currents are displayed in Figs.(\ref{fig:main_result}) (a) and (b). Both diagrams have a similar profile and are neatly separated into a conducting (yellow-red) and an 
insulating (light blue) region. The first occurs at $\Delta T,\Delta\beta>0$, where oscillator 1 has lower damping and higher temperature than oscillator 2
When $\Delta T,\Delta\beta<0$, the situation is reversed and the sistem is insulating. 

One can see here one remarkable feature: when $\Delta\beta$ is sufficiently negative ($\approx -0.02$), a \emph{negative} current flows at positive $\Delta T$. 
This means that, tuning the chemical potential, the system operates as a \emph{cooling machine} that pumps energy and spin from the colder to the hotter system. 

%Notice also that, if $\Delta\beta$ is sufficiently large ($\approx0.02$), the currents are positive at all gradients and the rectification effect is suppressed. 
%The amplification $A_M$ and $A_E$ of the currents with respect to their values
%at $\Delta T=\Delta\beta=0$ are shown in Figs.(\ref{fig:main_result}) (c) and (d).

Figure(\ref{fig:main_result}d) shows the power ratio $G=|p_1-p_2|/(p_1+p_2)$, which is roughly \emph{symmetric} in the $(\Delta T,\Delta\beta)$ plane. This means that, in both the conducting and the insulating region,
there is a similar difference in SW power. In the conducting region, the two oscillators are synchronized and there is a net energy and magnetization transfer from the "hot" to the "cold" system. On the contrary,
in the insulating region, there is no current and the energy provided by the baths is returned to them. 
This corresponds precisely to the situation described at the beginning of the paper: the condition $p_1\neq p_2$ \emph{is necessary, but not sufficient}, to have transport. 

An important parameter in spin-caloritronics is the spin-Seebeck coefficient, which describes the capability of the system to convert the energy current into
a spin current. However, this makes sense only in the linear regime, where the currents are proportional to che thermodynamic forces. 
Here, the performances of the system can be described by the current ratio $S=|j_M/j_E|$, which is displayed in Fig.(\ref{fig:main_result}d) in logarithmic scale. In the conducting region, one can see
that $S$ is higher in the quasi-linear regime (at small $\Delta T$ and high $\Delta\beta$), where it reaches the $60\%$, while it decreases smoothly until about $40-30\%$ as $\Delta T$ increases.
The current ratio drops to $13\%$ in the inversion regions, where the current becomes negative (resp. positive) at positive (resp. negative) gradient.  

\section{Comparison with micromagnetic simulations}
\label{sec:micro}

To check our model on a realistic system, we have performed micromagnetic simulations on a nano-pillar made of two Permalloy (Py) nano-disks, displayed in Fig. (\ref{fig:system})a. 
The disks have a radius $R=20$ nm, thicknesses $t_1=5$ and $t_2=3$ nm and are separated by a 4 nm spacer. An external field $\bm{H}_{\rm{ext}}=1$ T is applied along the $z$ direction. 
The other micromagnetic parameters of the system are The exchange stiffness of Py is $A=1.3\times 10^{-11}$ J/m. The magnetic parameters of the disks, 
taken from Ref.\cite{thesis}, are $M_{s1}=7.8\times 10^5$ A/m, $M_{s2}=9.4\times 10^5$ A/m, $\alpha_1=1.6\times 10^{-2}$, $\alpha_2=0.85\times 10^{-2}$ and $\gamma_0=1.87\times 10^{11}$ rad$\times$s$^{-1}$$\times$T$^{-1}$.
Those parameters are the same as in Refs.~[\onlinecite{thesis,borlenghi13}]. The computations were performed with the Nmag micromagnetic solver \cite{fischbacher07}, using a finite 
element tetrahedral mesh with a maximum size of 3 nm.

Starting from a uniform tilt of the magnetization of $8^\circ$ with respect to the $z$ direction, the time evolution was computed for 50 ns with a time-step of 1 ps,
and the results were averaged over 16 samples with different realization of the stochastic noise. 

\begin{figure}
\begin{center}
\includegraphics[width=8cm]{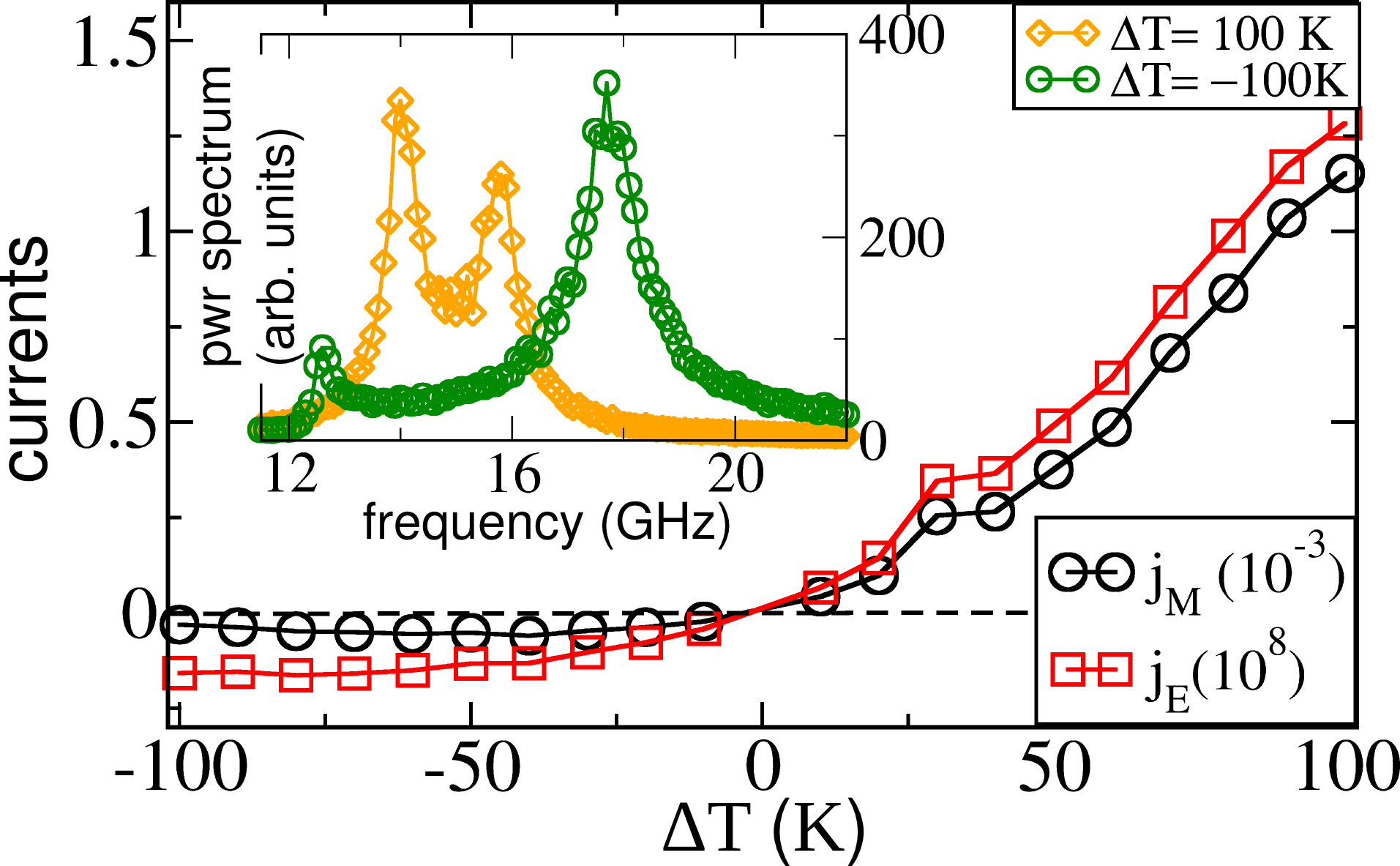}
\end{center}
\caption{(Color online). Time-averaged currents vs temperature difference in a spin-valve nano-pillar. The inset shows the overlap of the SW modes, at the core of the rectification effect.}
\label{fig:micromag}
\end{figure}

The time-averaged currents are shown in Fig.~\ref{fig:micromag} as a function
of the temperature difference between the two disks. Notice that the currents displayed here are per unit coupling, and are thus pure numbers. 

One can see that the system displays a strong rectification effect, 
(compare Fig.~\ref{fig:micromag} with Fig.~\ref{fig:rectification}). 
Moreover the SW spectra in conducting and insulating regimes are drastically different.
Indeed, for negative $\Delta T$ the SW spectra display two distinct maxima while 
for positive $\Delta T$ there is a single broadened peak (see the inset of Fig.~\ref{fig:micromag}). 
This picture is coherent with the simple double oscillator model
and suggests that the syncronization mechanism proposed above is indeed at the 
basis of the rectification observed in the realistic simulations of the nanopillar.

\section{Conclusions}

To conclude, we have studied through simple analytical arguments and computer simulations a novel system, which can rectify both energy and magnetization currents.
A significant rectification effect is present in a large set of system parameters and the underlying physical process suggests a new method for 
phase-locking and transfer of energy and magnetization in magnonic and spin-caloritronics devices. The connection with phase synchronization phenomena is insightful
and allows to understand the basic rectification mechanism in a simple way. 

We wish to stress that the results presented here are general and may apply to systems described by the Landau-Lifschiz-Gilbert (LLG) equation \cite{landau65,gilbert55,gurevich96}, 
with different geometries, coupling mechanisms and sizes between the nm and the $\mu$m range. The nonlinearity of the LLG equation and the presence of noise are the essential ingredients for this effect.
Chosing a spin-valve geometry allows to study a realistic system where ST plays also a significan role, controlling the magnon population of the device. However, we expect that devices with different geometries
(such as nano-contacts of different materials where the spins are echange-coupled) can exhibit a similar rectification effect.
At variance with the models studied in the context of phononics \cite{li12}, the magnonic device allows to consider 
coupled transport of the the two basic conserved quantities, energy and magnetization. The control of the 
associated forces allows for new possibilities. As exemplified in this work, it would be for instance
possible to control the energy current on the device scale by changing the applied spin-polarized currents. 

\begin{acknowledgments}
We gratefully acknowledge the Swedish Research Council (VR), Carl Tryggers foundation and G\"oran Gustafssons 
foundation for financial support. The computer simulations were performed on resources provided by the Swedish National Infrastructure for Computing (SNIC) at National Supercomputer Centre (NSC). 
\end{acknowledgments}

%\bibliography{biblio_spin_dimer}
%\bibliographystyle{plain}{99}
%\bibliographystyle{h-physrev}
%\bibliographystyle{unsrt}
\end{document}